\documentstyle[aps,prl,epsf,floats,twocolumn]{revtex}
\begin{document}
\draft \twocolumn[\hsize\textwidth\columnwidth\hsize\csname
@twocolumnfalse\endcsname
\title{Luttinger liquid superlattices}
\author{J. Silva-Valencia$^{\, (1)}\!$, E. Miranda$^{\, (1)}\!$,
and Raimundo R.\ dos Santos$^{\, (2)}$}
\address{$^{(1)}$Instituto de F\'{\i}sica ``Gleb Wataghin'', Unicamp, C.P. 6165,
13083-970 Campinas SP, Brazil\\
$^{(2)}$Instituto de F\' \i sica, Universidade Federal do Rio de Janeiro,
                 C.P.\ 68.528, 21945-970 Rio de Janeiro RJ, Brazil}
\date{\today}
\maketitle
\begin{abstract}
We calculate the correlation functions and the DC conductivity of
Luttinger liquid superlattices, modeled by a repeated pattern of
interacting and free Luttinger liquids. In a specific realization,
where the interacting subsystem is a Hubbard chain, the system
exhibits a rich phase diagram with four different phases: two metals
and two compressible insulators. In general, we find that the
effective low energy description amalgamates features of both types of
liquids {\em in proportion to their spatial extent}, suggesting the
interesting possibility of `engineered' Luttinger liquids.
\end{abstract}

]

In recent years, new experimental techniques have made it possible to
grow nanostructures which are topologically one-dimensional, such as
quantum wires and Carbon nanotubes.  However, care must be taken when
invoking existing models to discuss their electronic properties, since
inhomogeneities must be taken into account in a fundamental way.
Consider, for instance, the Luttinger liquid (LL), which is the
standard model for low-energy phenomena involving interacting
electrons in one dimension\cite{voit}.  The absence of conductance
renormalization in long high-mobility GaAs wires\cite{taru} has been
explained in terms of a usual LL (representing the wire) in contact
with a non-interacting LL at each of its ends (representing the Fermi
liquid leads); that is, overall, the system can be thought of as an
{\em inhomogeneous Luttinger Liquid} (ILL).  LL's with different
inhomogeneity profiles have also been used in the context of the
fractional quantum Hall effect (FQHE), to describe transitions between
edge states at different fillings\cite{oregf,chklo}, or between an
edge state connected to a Fermi liquid\cite{chamon}.  A different
class of inhomogeneous systems is represented by superlattices (SL's)
and multilayers.  By varying the relative thicknesses of the repeating
unit in magnetic metallic multilayers, fascinating properties such as
exchange oscillation and giant magneto-resistance (GMR) have been
found\cite{gmr}.  The interplay between electron correlations and a SL
structure (or layering) can therefore lead to collective properties
quite distinct from those of each of its constituents. Furthermore,
the ability to manipulate physical properties by choosing an
appropriate spatial modulation opens the way for a whole new set of
`engineered' materials.

With this in mind, our purpose here is to discuss the properties of a
one-dimensional SL made up of a periodic arrangement of two long and
perfectly connected LL's, one interacting and the other free.
Accordingly, the low-energy properties of this Luttinger liquid
superlattice (LLSL) are described by generalizing the usual bosonized
Hamiltonian\cite{voit,emery,hald} as follows
\begin{eqnarray}
\label{ham3}
H & = & \frac{1}{2\pi}\sum_{\nu =\rho ,\sigma }\int dx
\left\{ u_{\nu }(x)K_{\nu}(x)
 {\left( \partial_{x} \Theta_{\nu} \right)}^{2}
\right. \nonumber \\
 &   & \left. \mbox{} + \frac{u_{\nu }(x)}{K_{\nu}(x)}{\left(
\partial_{x}\Phi_{\nu}  \right)}^{2} \right\}
\end{eqnarray}
where the sum extends over separated charge- ($\nu=\rho$) and spin-
($\nu=\sigma$) degrees of freedom, each of which with layer-dependent
parameters $u_{\nu }(x)$ and $K_{\nu }(x)$; these determine,
respectively, the velocity of elementary excitations and the algebraic
decay of correlations functions for each degree of freedom.  For $x$
on the free layer one has $K_\nu(x)=1$ and $u_\nu(x)=v_{F}$, the Fermi
velocity, whereas for $x$ on the repulsive layer $K_\nu(x)$ and
$u_\nu(x)$ become the usual uniform LL parameters.  For definiteness,
we will often speak of a `Hubbard superlattice' (HSL), where the
interacting layer is taken as a Hubbard model with hopping $t$ and
on-site repulsion $U$\cite{rrds1,rrds3}; a weak coupling perturbation
theory similar to that of the homogeneous model can be used to show
that Eq.\ (\ref{ham3}) indeed describes the low energy and small
momentum sector of the discrete model with long layers \cite{safi3}.
In the homogeneous case, the dependence of the LL parameters on both
the density and $U$ has been determined by recourse to the exact
solution\cite{schulz,liebwu,frahm,yang}.  With respect to magnetic
properties, the SL structure does not break spin SU(2) symmetry, so
that the inhomogeneous $K_{\sigma }$ is still expected to renormalize
to $ K_{\sigma }^{\ast }=1$.  The spin sector stiffness is therefore
unrenormalized as in the homogeneous system\cite{odd-even}.

The boson phase fields $\Phi_{\nu}$ are related to the charge and spin
densities, $\rho $ and $\sigma $, through $\sqrt{2}\partial
_{x}\Phi_{\nu}(x)/\pi =\nu $, while $\Theta_{\nu}$ is such that $
\partial _{x}\Theta _{\nu }$ is the momentum field conjugate to
$\Phi _{\nu } $: $[\Phi _{\nu}(x),\partial _{y}\Theta _{\nu^\prime
}(y)]= i\delta _{\nu ,\nu^\prime}\delta (x-y)$. $\Phi_{\nu}$ and
$\Theta_{\nu}$ are dual fields, since they satisfy both
\begin{equation}
\partial_{t}\Phi_{\nu}=u_{\nu}(x)K_{\nu}(x)\partial_{x}\Theta_{\nu}
\label{em1}
\end{equation}
and the equation obtained through the replacements
$\Phi_\nu\to\Theta_\nu$,
$\Theta_\nu\to\Phi_\nu$, and
$K_\nu\to 1/K_\nu$.
These equations can be uncoupled to yield
\begin{equation}
\partial _{tt}\Phi _{\nu} - u _{\nu} K _{\nu} \partial _{x}
\left(\frac{u _{\nu}}{K _{\nu}} \partial _{x}\Phi _{\nu}\right)=0,
\label{emphi}
\end{equation}
and a dual equation for $\Theta_\nu$.

The equations of motion are subject to the continuity of $\Phi _{\nu}$
and $\Theta _{\nu}$\cite{safi1,safi2} (which ensures the continuity of
the fermionic field).  Since their time derivatives are also
continuous, Eq.~(\ref{em1}) and its dual yield, as additional
conditions, the continuity of $\left(u_{\nu
}/K_\nu\right)\partial_{x}\Phi_\nu$ and $u_\nu
K_\nu\partial_{x}\Theta_\nu$ at the contacts.  Thus, both charge and
spin currents $j_{\nu }=\sqrt{2}\partial _{t}\Phi _{\nu }/\pi $ are
conserved, since we neglect Umklapp processes and spin backscattering.

The Hamiltonian (\ref{ham3}) is straightforwardly diagonalized by a
normal mode expansion
\begin{eqnarray}
\Phi _{\nu }(x,t)& = & -i\sum_{p\neq 0}{\rm sgn}(p)
\frac{\phi _{p,\nu }(x)}{2\sqrt{\omega_{p,\nu }}}
[b_{-p,\nu }e^{i\omega_{p,\nu }t}+b_{p,\nu }^{\dagger }e^{-i\omega_{p,\nu }t}]
\nonumber \\
\label{phi}
 &   & -\phi _{0,\nu }(x)+ \gamma_{\lambda \nu}t\\
\Theta_{\nu }(x,t)& = & i\sum_{p\neq 0}
\frac{\theta _{p,\nu }(x)}{2\sqrt{\omega_{p,\nu }}}
[b_{-p,\nu }e^{i\omega_{p,\nu }t}-b_{p,\nu }^{\dagger }e^{-i\omega_{p,\nu }t}]
\nonumber \\
 & & +\theta _{0,\nu }(x)- \tau_{\lambda \nu}t\label{thet}
\end{eqnarray}
where $b_{p,\nu}^{\dagger}$ are boson creation operators ($p>0$).
$\phi_{0,\nu }(x)$ and $\theta_{0,\nu}(x)$ are the zero mode functions
which, in the homogeneous case, are given by $\phi _{0,\nu
}(x)=N_{\nu}\frac{\pi x}{L}$ , $\theta _{0,\nu }(x)=J_{\nu}\frac{\pi
x}{L}$, where $N_{\nu}$ and $J_{\nu}$ are the (charge and spin) number
and current operators.  Besides, in this case $\gamma_{\nu}=\pi
u_{\nu}K_{\nu} J_{\nu}/L$ and $\tau_{\nu}=\pi
(u_{\nu}/K_{\nu})N_{\nu}/L$. However, in a LLSL the inhomogeneity will
induce a modulation of the charge (but not of the spin) density of the
system.  Thus, one needs to introduce in general {\em layer-specific}
number and current operators.  Since each layer is a LL, the
variations across it are $\Delta\phi_{0,\nu}=\pi N_{\lambda
\nu}$ and $\Delta\theta_{0,\nu}=\pi J_{\lambda \nu}$, where
$\lambda=0$ or $U$, depending on whether it is a free or interacting
layer, respectively. $\phi_{0,\nu }(x)$ and $\theta_{0,\nu}(x)$ will
then be linear continuous functions of $x$, with slopes given by the
layer number and current operators (we omit the expressions for
brevity).  Analogously, from the equations of motion (\ref{em1}), we
obtain $\gamma_{\lambda\nu}$ and $\tau_{\lambda\nu}$.

In order to find the equilibrium value of the density in each layer,
one needs to equate their chemical potentials
\begin{equation}
\mu_{0}(n_0)=\mu_{0}\left( n+\ell (n-n_{U})\right)=\mu_U(n_U),
\label{equi}
\end{equation}
where $n=N/L$ is the total electron density, $\ell=L_U/L_0$ and
$\mu_\lambda$ and $n_\lambda$ are the chemical potential and density
of each layer, respectively.  For definiteness, we have determined the
charge profile in a HSL using the exact expression for
$\mu_U(n_U)$\cite{liebwu}.  We found that the charge tends to
accumulate in the free layer.  This is rather intuitive, since
electrons decrease their mutual repulsion energy by flowing into the
free layer. This was observed in numerical studies of the
HSL\cite{rrds3}. Of course, such a charge inhomogeneity will be
strongly suppressed with the inclusion of long-range Coulomb
interactions, which are absent in a Hubbard model description.

\begin{figure}[htbp]
\epsfxsize=3.5in \epsfbox{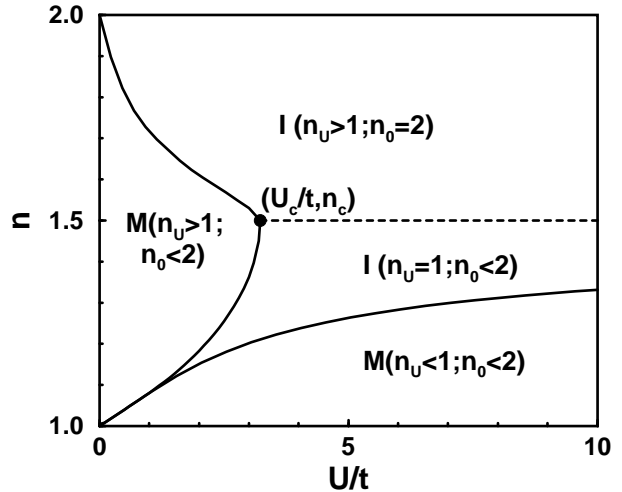}
\caption{Phase diagram of a Hubbard superlattice showing two metallic
(M) and two insulating (I) phases ($\ell=1$) ($U_c/t\approx 3.2309$
and $n_c=(2+\ell)/(1+\ell)$).  }
\label{phase}
\end{figure}

The HSL has a very rich phase diagram. For $n<1$ the system is always
metallic. For $n>1$, however, we observe four different phases, two
metallic and two insulating, each characterized by its charge profile,
as shown in Fig.~\ref{phase}. The two insulating phases correspond to
either $n_U=1$ (Mott insulator) or $n_0=2$ (band insulator). Note
that, in each case, one type of layer is by itself insulating while
the other is metallic, the overall insulating character being a
consequence of the 1D structure (`resistors in series'). Therefore,
both insulating phases are {\em gapless}, except at the phase boundary
indicated by the dashed line in Fig.~\ref{phase}, where {\em both}
$n_U=1$ {\em and} $n_0=2$ and the system exhibits a Mott-Hubbard
gap. The density at this line is thus
$n_c=(2+\ell)/(1+\ell)$\cite{rrds2}.  The two metallic phases differ
in the density of the interacting layer: the chemical potential can
fall in either its upper or its lower Hubbard band. For values of
$U>U_c\approx 3.2309 t$, the system cannot sustain the upper metallic
phase. This value is independent of the `aspect ratio' $\ell$. Note
that, for $U<U_c$, the HSL is always gapless.

While the topology of the phase diagram of Fig.~\ref{phase} is
specific to a HSL, we expect that several of its features should be
generic to other LLSL's. In particular, the `division of labor'
between the two subsystems, where one is responsible for the
insulating behavior whereas the other renders the system compressible,
is reflected in the weighted form of the SL compressibility
\begin{equation}
\kappa_{s}=\frac{1}{L}\left(\frac{\partial^{2}E_{0}}{\partial
N^{2} }\right)^{-1}=\frac{\kappa_0 + \ell\kappa_{U}}{1+\ell},
\label{comsu}
\end{equation}
where $\kappa_U = 2K_\rho/\pi u_\rho$ and $\kappa_0 = 2/\pi v_F$ are
the compressibilities of the interacting and free layers respectively.

The SL structure also affects the velocity of excitations.  For $p\ll
\pi/(L_{U}+L_{0})$, the dispersion relation of the LLSL is linear,
with effective velocities
\begin{equation}
c_{\nu }=\frac{v_{F}(1+\ell)}
{\sqrt{1+\Delta_\nu \ell v_F/u_\nu + \left(\ell v_F/u_\nu\right)^2}},
\label{cnu}
\end{equation}
where $\Delta _{\nu }=K_{\nu }+K_\nu^{-1}$.  Clearly, $c_\nu\to u_\nu$
as $\ell\to\infty$, and $c_\nu\to v_F$ as $\ell\to 0$.  Furthermore,
as one approaches the insulating phase from the low-density region
(see Fig. \ref{phase}), $c_\rho\to 0$ as a result of $u_\rho\to 0$ in
the interacting layer.  As in the homogeneous system, the LL
description breaks down whenever a gap opens in the charge or spin
sector of either layer. In the HSL case, this happens in both
insulating regions of Fig.~\ref{phase}. Note, however, that the
determination of the phases through Eq.~(\ref{equi}) does not rely on
the LL description.

We now focus on the correlations.  The $T=0$ asymptotic behavior
(i.e., for well separated $x$ and $y$) of charge and spin correlations
is given by
\begin{eqnarray}
\left\langle n(x)n(y)\right\rangle  &\sim &\frac{\alpha _{\rho }}{\pi
^{2}\left| x-y\right| ^{2}}+A_{1}\frac{e^{2i\left(
\overline{\phi}_{0}(y)-\overline{\phi}_{0}(x)\right) }}
{\left| x-y\right| ^{1+K_{\rho }^{\ast }}}
\nonumber \\
&&+A_{2}\frac{e^{4i\left( \overline{\phi}_{0}(y)-
\overline{\phi}_{0}(x)\right) }}{\left| x-y\right| ^{4K_{\rho }^{\ast }}}, \\
\left\langle {\bf S}(x).{\bf S}(y)\right\rangle &\sim& \frac{\alpha
_{\sigma }}{ \pi ^{2}\left| x-y\right|
^{2}}+B_{1}\frac{e^{2i\left( \overline{\phi}_{0}(y)-
\overline{\phi}_{0}(x)\right) }}{\left| x-y\right| ^{1+K_{\rho }^{\ast }}},
\end{eqnarray}
where $\overline{\phi}_{0}(x)=k_F x-\phi_{0,\rho}(x)$ and the LLSL 
effective exponent is
\begin{equation}
K_{\rho }^{\ast }=\frac{\sqrt{1+\Delta _{\rho }\ell v_F/u_\rho
+\left(\ell v_F/u_\rho\right) ^{2}}}{1+\ell v_F/K_\rho u_\rho}
\equiv f(K_\rho),
\label{krhostar}
\end{equation}
here $\alpha_{\nu}$ is a function of system parameters and the layer.
Similarly, correlation functions for singlet and triplet
superconducting pairing are
\begin{equation}\label{STP}
 \left\langle O^{\dagger}(x)O(y)\right\rangle\sim
 \frac{C}{\left| x-y\right| ^{1+\overline{K}_{\rho }}},
\end{equation}
where $\overline{K}_{\rho }=f(1/K_\rho)$.  One should note that the
correlation functions depend not only on the difference $x-y$, but
also on the actual positions $x$ and $y$, through the zero mode
functions.  Their effect will be to generate the usual spatial
oscillations present in homogeneous LL's.  However, due to the
inhomogeneous density profile, their period will vary from layer to
layer, reflecting the layer-dependent Fermi wave-vectors; this is akin
to the oscillatory behavior of the exchange coupling in magnetic
metallic multilayers\cite{gmr,rrds3}.  In spite of the presence of
effective exponents $K_\rho^\ast$ and $\overline{K}_\rho$, the
condition for dominant superconducting correlations reduces to the one
for homogeneous systems, namely $K_\rho>1$.  The dominant term in the
charge and spin correlation functions is the second one, which in the
homogeneous case corresponds to the $2k_{F}$ contribution. This
predominance, however, may be superseded by the behavior of the
amplitude $A_1$, as discussed in Ref.\ \cite{cdw}.

\begin{figure}[tbp]
\epsfxsize=3.5in \epsfbox{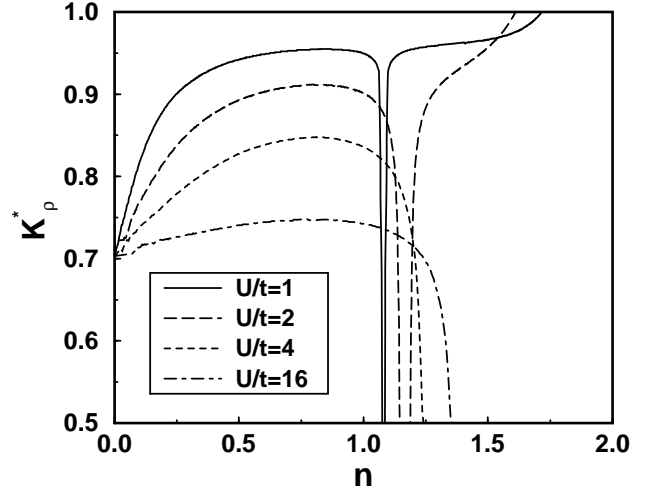}
\caption{The correlation exponent $K_{\rho }^{\ast}$ of a Hubbard
superlattice as a function of the total electron density $n$ for
$\ell=1$ and different values of $U$.  }
\label{Krhon}
\end{figure}

In Fig.~\ref{Krhon}, the correlation exponent $K_{\rho }^{\ast}$ of a
HSL is shown as a function of filling, for $\ell = 1$.  Both metallic
phases are characterized by $1/2<K_{\rho }^{\ast}<1$.  On the low
density side, $K_{\rho }^{\ast}$ approaches a value larger than 1/2 as
$n\to 0$, which depends on $\ell$ but not on $U$. This is a feature
unique to the LLSL. Note from Eq.~(\ref{krhostar}) that $K_{\rho
}^{\ast}$ interpolates monotonically between 1 (the free value) and
$K_{\rho}$ (the interacting layer exponent) as $\ell$ is varied from
$0$ to $\infty$. This illustrates a general feature of the LLSL,
namely: {\em by varying the `aspect ratio' $\ell$, one can fine-tune a
physical property to a specified value}.

Finally, we discuss the transport properties of a LLSL.  In the
presence of an applied electric field the equation of motion for
$\Phi_{\rho}$ becomes\cite{safi1,safi2}
\begin{equation}
\left[-\frac{\partial_{tt}}{u_{\rho}K_{\rho}}+
\partial_{x}\left(\frac{u_{\rho}}{K_{\rho}}\partial_{x}\right)\right]
\Phi_{\rho}(x,t)=-eE(x,t).
\end{equation}
The nonlocal conductivity is given by
\begin{equation}
\sigma (x,y,t)=-\frac{2g_{o}}{\pi }\ \partial _{t}G(x,y,t)
\label{sigma}
\end{equation}
where $g_{o}=e^2/h$ is the conductance quantum and $G(x,y,t)=-i\
\theta (t)\left\langle \left[ \Phi _{\rho} (x,t),\Phi _{\rho}
(y,0)\right] \right\rangle$ is the bosonic Green's function.

We first calculate the Drude conductivity, which gives the current
response to a uniform electric field: $\lim_{\omega\to 0}
\sigma(q=0,\omega)$ \cite{fenton}. A straightforward calculation yields
\begin{equation}
\sigma(q=0,\omega)=2g_oc_\rho K_\rho^\ast\ \delta(\omega ),
\label{drude}
\end{equation}
The delta function coefficient is the Drude weight.  It has the same
form as for the homogeneous case\cite{schulz}, but with the effective
velocity and effective exponent replacing the corresponding uniform
quantities $u_\rho$ and $K_\rho$. By plugging in the results from
Eqs.~(\ref{cnu}) and (\ref{krhostar}), one recognizes the conductivity
of resistors connected in series.

\begin{figure}[tbp]
\epsfxsize=3.5in \epsfbox{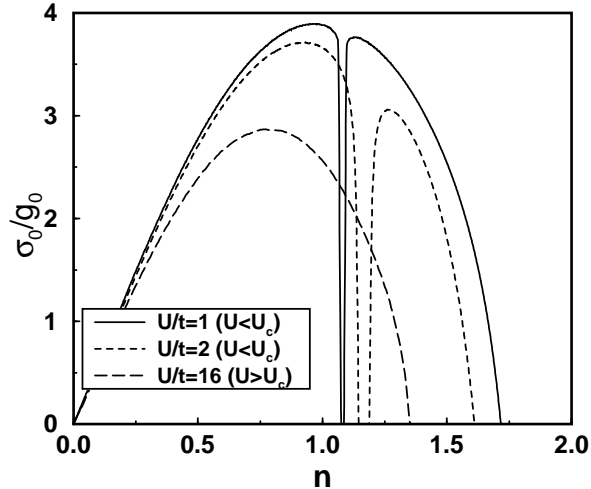}
\caption{The Drude weight  of a Hubbard superlattice as a function  of
$n$ for $\ell=1$ and different values of $U$.}
\label{drun} 
\end{figure}

We have plotted the Drude weight of a HSL as a function of $n$ for
$\ell=1$ and several values of $U$ in Fig.~\ref{drun}. The plot shows
the re-entrant behavior as a function of $n$ for $U<U_c$. Furthermore,
the Drude weight dips to zero upon approaching the insulating regions
as a result of the vanishing charge velocities $u_\rho\to 0$ (Mott
insulator) and $v_F \to 0$ (band insulator).

A more common experimental situation occurs when a field is applied to
a finite region of the sample. In this case, the inverse order of
limits applies $\lim_{q\to 0} \sigma(q,\omega=0)$, and we obtain the
Landauer conductance \cite{fenton}. In the LLSL
\begin{equation}
\sigma(q,\omega=0)=2g_oK_{\rho}^\ast\ \delta(q).
\label{landauer}
\end{equation}
Once again, the result is in close analogy with the homogeneous system
\cite{apel}, where the SL interaction exponent has replaced the
homogeneous one. Both Drude and Landauer responses, therefore, can be
modulated by changing the SL spacing. However, the interaction
renormalization of Eq.~(\ref{landauer}) is not revealed in usual DC
conductance measurements, where it is masked by the presence of the
Fermi liquid leads \cite{safi1}. In a LLSL of length $L$, only by
going to frequencies of the order of the inverse traversal time
$\omega>u_\rho/L$ can the influence of the $K_\rho^\ast$ exponent be
felt \cite{matveev}.

In closing, we would like to highlight the fact that the low energy,
long wavelength properties of a LLSL can be, in effect, subsumed into
a few effective parameters, in close analogy with the usual
homogeneous LL description. These effective parameters, on the other
hand, turn out to be weighted averages of the underlying subsystem
properties, in rough proportion to their spatial extent [see, e.g.,
Eqs.~(\ref{comsu}), (\ref{cnu}) and (\ref{krhostar})].  Such a
`tempered Luttinger liquid' description suggests the interesting
possibility of creating SL structures with properties engineered to
suit a particular purpose, in a way reminiscent of modulation-doped
semiconductor heterostructures and magnetic multilayers. Whether this
will prove feasible, however, remains to be seen.

In summary, we have considered Luttinger liquid superlattices made up
of a periodic arrangement of free and repulsive Luttinger liquids.
Due to the space-dependent properties of the system, a non-homogeneous
charge profile ensues. A specific realization of such a system, a
Hubbard superlattice, was investigated in detail and its phase diagram
was shown to exhibit two metallic phases and two peculiar compressible
insulating ones.

The authors are grateful to A O Caldeira and T Paiva for discussions. 
Financial support from the Brazilian Agencies CNPq, FAPESP (E.M.), and
FAPERJ (R.R.d.S.) is also gratefully acknowledged.

\end{document}